\documentclass{iau307}
\usepackage{graphicx}
\usepackage{natbib}
\usepackage{url}
\usepackage{dtklogos}
\bibpunct{(}{)}{;}{a}{}{,}

\title[Abundance analysis of HD~22920 spectra] 
{Abundance analysis of HD~22920 spectra\protect\thanks{Based on observations obtained at the Canada-France-Hawaii Telescope (CFHT) which is operated by the National Research Council of Canada, the Institut National des Sciences de l'Univers of the Centre National de la Recherche Scientifique of France, and the University of Hawaii.}
}

\author[V. Khalack \& P. Poitras]   
{Viktor Khalack$^1$
 \and Patrick Poitras$^1$}

\affiliation{$^1$ Universit\'{e} de Moncton, Moncton, Canada \\ email: {\tt khalakv@umoncton.ca} \\[\affilskip]}

\pubyear{2014}
\volume{307}
\pagerange{}
\jname{New windows on massive stars: asteroseismology, interferometry, and spectropolarimetry}
\editors{G. Meynet, C. Georgy, J.H. Groh \& Ph. Stee, eds.}

\begin{document}

\maketitle

\begin{abstract}
The new spectropolarimetric observations of HD~22920 with ESPaDOnS at CFHT reveal a strong variability of its spectral line profiles with the phase of stellar rotation. We
have obtained $T_{\rm eff}$ = 13640 K, $\log{g}$=3.72 for this star from the best fit of its nine Balmer line profiles. The respective model of stellar atmosphere was calculated to perform abundance analysis of HD~22920 using the spectra obtained for three different phases of stellar rotation. We have found that silicon and chromium abundances appear to be
vertically stratified in the atmosphere of HD~22920. Meanwhile, silicon shows hints for a possible variability of vertical abundance stratification with rotational phase.
\end{abstract}

\firstsection 
\section{Introduction}

Accumulation or depletion of chemical elements at certain optical depths brought about by atomic diffusion can modify the structure of stellar atmospheres and it is therefore important to gauge the intensity of such stratification. Recently, we have obtained three ESPaDOnS spectra of magnetic CP star HD~22920 with the aim to study vertical stratification of chemical species in its stellar atmosphere. This star shows a weak photometric variability with a period P=3$^d$.95 \citep{Bartholdy88}. 
Its small value of Vsin(i)=30 km/s
results in comparatively narrow and unblended line profiles which are suitable for abundance analysis. The slow rotation and the presence of a weak magnetic field support the hypotheses of a hydrodynamically stable atmosphere, which is necessary for diffusion to take place.


\section{Spectral analysis\label{analysis}}

The preliminary results of the abundance analysis are presented here for HD~22920. The line profile simulation is performed using the ZEEMAN2 spectrum synthesis code \citep{Landstreet82} 
and LTE stellar atmosphere model calculated with PHOENIX \citep{Hauschildt+97} 
for the $T_{\rm eff}$ = 13640 K, $\log{g}$=3.72. 
For each element we have selected a sample of unblended lines, which are clearly visible in the analyzed spectra. The element's abundance, radial velocity and Vsin(i) were fitted using an automatic minimization routine independently for each line profile in each rotational phase, and are presented in Table~\ref{tab1}. For each ion presented in the Table~\ref{tab1}, the number of analyzed lines is specified in brackets. Our estimates of the radial velocity and $V \sin{i}$ are consistent with the previously published results for this star.

\begin{figure}[t]
\begin{center}
\includegraphics[width=2.1in,angle=-90]{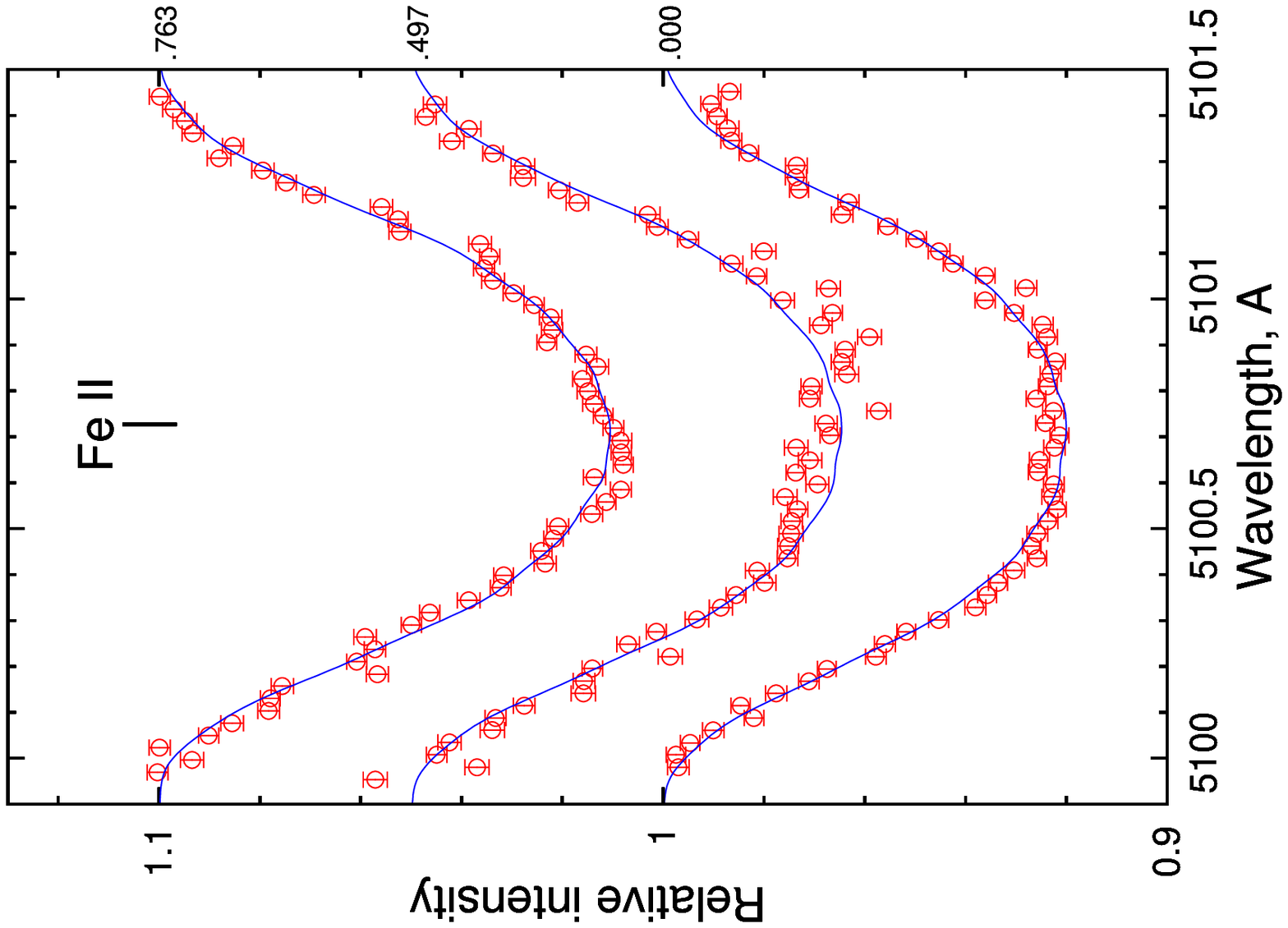}
\includegraphics[width=2.1in,angle=-90]{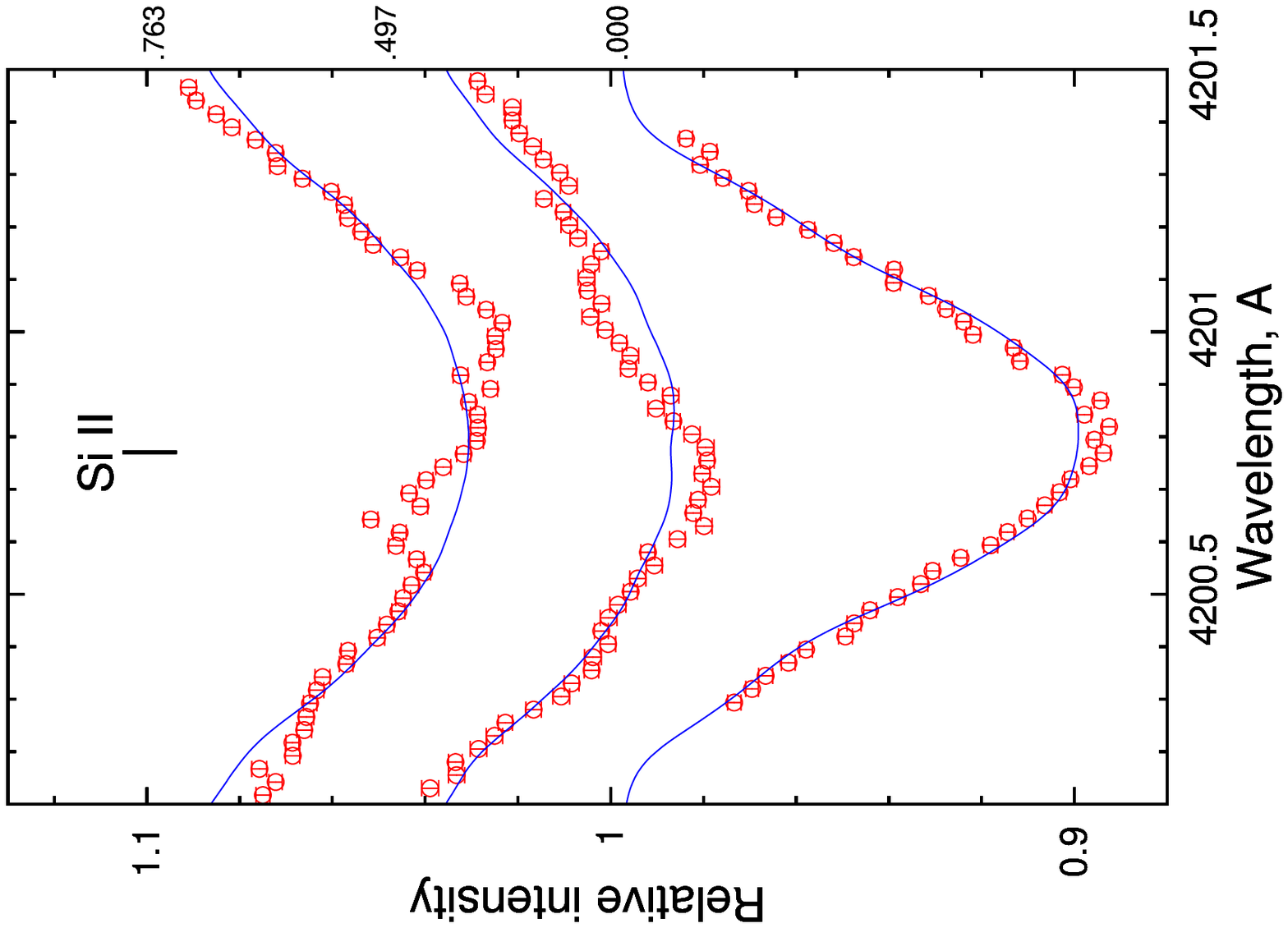}
\includegraphics[width=2.1in,angle=-90]{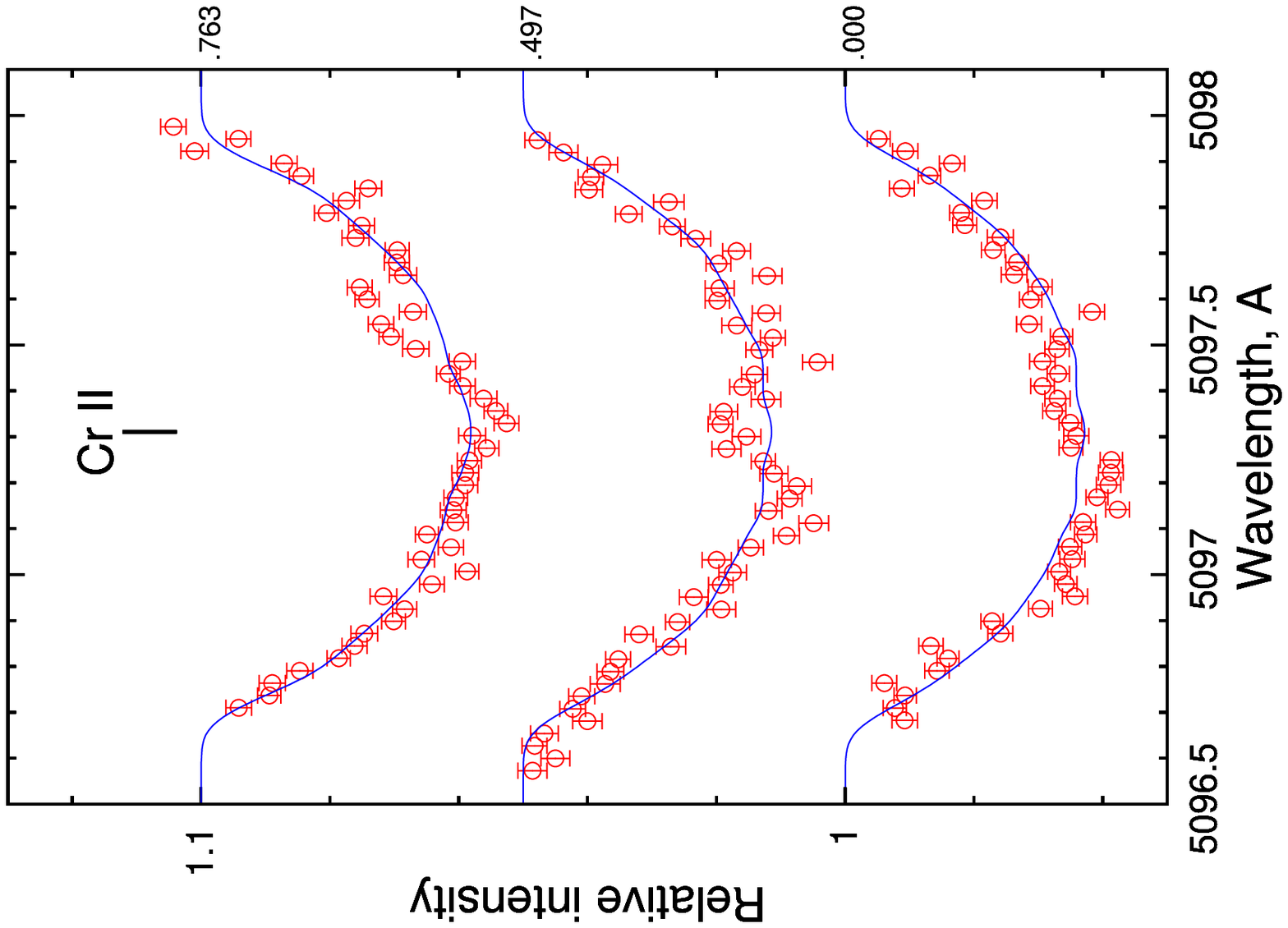}
\caption{The best fit (continuous line) obtained for the observed FeII 5100\AA, SiII 4200\AA\, and CrII 5097\AA\, line profiles (open circles) in HD~22920 at different phases of stellar rotation.}
\label{fig1}
\end{center}
\end{figure}

For each rotational phase we have determined average abundance of oxygen, silicon, iron and chromium 
(see Table~\ref{tab1}). During simulation of the synthetic line profiles we assume a homogeneous horizontal distribution of elements abundance, which is not correct, but provides an estimate for the average abundance for the given rotational phases. The observed profiles (see Fig.~\ref{fig1}) differ significantly from the simulated ones. This and the obvious variability of line profiles with rotational phase argue in favor of a non-homogeneous horizontal abundance distribution for these elements. Among the studied elements, SiII line profiles show the strongest variability with rotational phase. 

\begin{table}
\begin{center}
\caption{Average abundance of chemical species in the atmosphere of HD~22920 for different rotational phases. Number in parenthesis represents the number of analyzed lines for each ion.}
\label{tab1}
\begin{tabular}{l|cr|cr|cr}\hline
\textbf{Abundance} & \multicolumn{2}{c|}{\textbf{$\varphi$ =0.0}} & \multicolumn{2}{c|}{\textbf{$\varphi$=0.497}} & \multicolumn{2}{c}{\textbf{$\varphi$ =0.763}} \\
\hline
$\log(N({\rm OI})/N_{tot})$    &	-4.03$\pm$0.18 &(2)	 & -4.11$\pm$0.20  &(3) & -4.07$\pm$0.18 & (3) \\
$\log(N({\rm SiII})/N_{tot})$  &	-3.88$\pm$0.38 &(15) & -4.34$\pm$0.29  &(8) & -4.20$\pm$0.29 &(10) \\
$\log(N({\rm FeII})/N_{tot})$  &	-4.26$\pm$0.22 &(32) & -4.11$\pm$0.24 &(31) & -4.05$\pm$0.27 &(35) \\
$\log(N({\rm CrII})/N_{tot})$  &	-5.83$\pm$0.95 &(6)  & -5.59$\pm$1.31 & (5) & -5.27$\pm$1.18  &(5) \\
$V \sin{i}$ (km/s) &  36.1$\pm$4.3 & (55) &  37.8$\pm$5.4 & (47) & 37.3$\pm$5.0 & (53) \\
$V_{\rm r}$ (km/s) &  16.1$\pm$4.0 & (55) &  18.6$\pm$4.5 & (47) & 19.8$\pm$4.4 & (53) \\
\hline
\end{tabular}
\end{center}
\end{table}

\section{Summary\label{summary}}
Here we presented our estimate of $T_{\rm eff}$ = 13640K, $\log{g}$ =3.72 in HD~22920 and the average abundances of oxygen, silicon, iron and chromium for different rotational phases (see Table~\ref{tab1}). All the analyzed elements show variability of their line profiles with rotational phase and seem to have non-uniform horizontal distributions of their abundance. Among the studied elements only silicon and chromium abundances appear to be vertically stratified in the stellar atmosphere of HD~22920.

\bibliographystyle{iau307}
\bibliography{KhalackP2}







\end{document}